\documentclass[preprint,showpacs,showkeys,preprintnumbers,amsmath,amssymb]{revtex4}

\usepackage{graphicx,epsfig}% Include figure files

\usepackage{dcolumn}% Align table columns on decimal point

\usepackage{bm}% bold math

\def\ut#1{\mathop{\vtop{\ialign{##\crcr

     $\hfil\displaystyle{#1}\hfil$\crcr\noalign

     {\kern1pt\nointerlineskip}\hbox{$\hfil\sim\hfil$}\crcr

     \noalign{\kern1pt}}}}}

\begin{document}

\preprint{}

\title{Apoastron Shift Constraints on Dark Matter Distribution at the Galactic Center}% Force line breaks

%with \\

\author{A.F. Zakharov}

\email{zakharov@itep.ru}

\affiliation{National Astronomical Observatories of Chinese
Academy of Sciences, Beijing 100012, China} \altaffiliation[Also
at~]{Institute of
Theoretical and Experimental Physics, Moscow, 117259, Russia;\\
CAMP, %\\
National University of Sciences and Technology, Rawalpindi, Pakistan;\\
BLTP, Joint Institute for Nuclear Research, Dubna,   Russia
%Lines break
%automatically or can be forced with \\
}
\author{A.A. Nucita}%
 \email{anucita@sciops.esa.int}
%\altaffiliation[Now at~]{XMM-Newton Science Operations Centre,
%ESAC, ESA, PO Box 50727, 28080 Madrid, Spain }
\affiliation{Department of Physics and {\it INFN}, University of
Lecce, CP 193, I-73100 Lecce, Italy;\\
XMM-Newton Science Operations Centre, ESAC, ESA, PO Box
50727, 28080 Madrid, Spain} %\altaffiliation[Now at~]{XMM-Newton
%Science Operations Centre, ESAC, ESA, PO Box 50727, 28080 Madrid,
%Spain }\emph{}
\author{F. De Paolis}%
\email{Francesco.DePaolis@le.infn.it} \affiliation{Department of
Physics and {\it INFN}, University of Lecce, CP 193, I-73100
Lecce, Italy}
\author{G. Ingrosso}
\email{Gabriele.Ingrosso@le.infn.it}
\affiliation{Department of Physics and {\it INFN}, University of
Lecce, CP 193, I-73100 Lecce, Italy}

\date{\today}% It is always \today, today,


\begin{abstract}

The existence of dark matter (DM) at scales of few pc down to
$\simeq 10^{-5}$ pc around  the centers of galaxies and in
particular in the Galactic Center region has been considered in
the literature. Under the assumption that such a DM clump,
principally constituted by non-baryonic matter (like WIMPs) does
exist at the center of our galaxy, the study of the $\gamma$-ray
emission from the Galactic Center region allows us to constrain
both the mass and the size of this DM sphere. Further constraints
on the DM distribution parameters may be derived by observations
of bright infrared stars around the Galactic Center. Hall and
Gondolo \cite{hallgondolo} used estimates of the enclosed mass
obtained in various ways and tabulated by Ghez et al.
\cite{Ghez_2003,Ghez_2005}. Moreover, if a DM cusp does exist
around the Galactic Center it could modify the trajectories of
stars moving around it in a sensible way depending on the DM mass
distribution. Here, we discuss the constraints that can be
obtained with the orbit analysis of stars (as S2 and S16) moving
inside the DM concentration with present and next generations of
large telescopes. In particular, consideration of the S2 star
apoastron shift may allow improving limits on the DM mass and
size.
\end{abstract}
%Use showkeys class option if keyword
\maketitle
\section{\label{intro} Introduction }%\lowercase{via} \textbackslash\textbackslash}

%\section{\label{sec:level1}First-level heading:\protect\\ The line

%break was forced \lowercase{via} \textbackslash\textbackslash}

In the last years intensive searches for dark matter (DM),
especially its non-baryonic component, both in galactic halos and
at galaxy centers have been undertaken (see for example
\cite{Bertone_2005,Bertone_2005a} for recent results). It is
generally accepted that the most promising candidate for the DM
non-baryonic component is neutralino. In this case, the
$\gamma$-flux from galactic halos (and from our Galactic halo in
particular) could be explained by neutralino annihilation
\citep{Gurevich_1997,Bergstrom_1998,Tasitsiomi_2002,Stoehr_2003,Prada_2004,Prada_2004a,Profumo_2005,Mambrini_2005}.
Since $\gamma$-rays are detected not only from high galactic
latitude, but also from the Galactic Center, there is a wide
spread hypothesis (see \cite{Evans_2004} for a discussion) that a
DM concentration might be present at the Galactic Center. In this
case the Galactic Center could be a strong source of $\gamma$-rays
and neutrinos
\citep{Bouquet_1989,Stecker_1988,Berezinsky_1994,Bergstrom_1998,Bertone_2004,Gnedin_2004,Bergstrom_2005,Horns_2005,Bertone_2005,Bertone_2005b}
due to DM annihilation. Since it is also expected that DM forms
spikes at galaxy centers
\citep{Gondolo_1999,Ullio_2001,Merritt_2003} the $\gamma$-ray flux
from the Galactic Center should increase significantly in that
case.

At the same time, progress in monitoring bright stars near the
Galactic Center have been reached recently
\citep{Genzel_2003,Ghez_2003,Ghez_2005}. The astrometric limit for
bright stellar sources near the Galactic Center with 10 meter
telescopes is today $\delta \theta_{10} \sim 1$~mas and the Next
Generation Large Telescope (NGLT) will be able to improve this
number at least down to $\delta \theta_{30} \sim 0.5$~mas
\citep{Weinberg_2005,Weinberg_2005b} or even to $\delta
\theta_{30} \sim 0.1$~mas
\citep{Report_2002,Weinberg_2005,Weinberg_2005b} in the K-band.
Therefore, it will be possible to measure the proper motion for
about $\sim 100$ stars with astrometric errors several times
smaller than errors in current observations.

The aim of this paper is to constrain the parameters of the DM
distribution possible present around the Galactic Center by
considering the induced apoastron shift due to the presence of
this DM sphere and either available data obtained with the present
generation of telescopes (the so called {\it conservative} limit)
and also expectations from future NGLT observations or with other
advanced observational facilities.

\section{The mass concentration at the Galactic Center}

Recent advancements in infrared astronomy are allowing to test the
scale of the mass profile at the center of our galaxy down to tens
of AU. With the Keck 10 m telescope, the proper motion of several
stars orbiting the Galactic Center black hole have been monitored
and almost entire orbits, as for example that of the S2 star, have
been measured allowing an unprecedent description of the Galactic
Center region. Measurements of the amount of mass $M(<r)$
contained within a distance $r$ from the Galactic Center are
continuously improved as more precise data are collected. Recent
observations \citep{Ghez_2003} extend down to the periastron
distance ($\simeq 3\times 10^{-4}$ pc) of the S16 star and they
correspond to a value of the enclosed mass within $\simeq 3\times
10^{-4}$ pc of $\simeq 3.67\times 10^{6}$ M$_{\odot}$. Several
authors have used these observations to model the Galactic Center
mass concentration. Here and in the following, we use the three
component model for the central region of our galaxy based on
estimates of enclosed mass given by Ghez et al
\cite{Ghez_2003,Ghez_2005} recently proposed \cite{hallgondolo}.
This model is constituted by the central black hole, the central
stellar cluster and the DM sphere (made of WIMPs), i.e.
\begin{equation}
M(<r)=M_{BH}+M_*(<r)+M_{DM}(<r)~, \label{totalmass}
\end{equation}
where $M_{BH}$ is the mass of the central black hole Sagittarius
A$^*$. For the central stellar cluster, the empirical mass profile
is
\begin{equation}
M_*(<r)=\left\{
\begin{array}{ll}
M_*\left(\frac{r}{R_*}\right)^{1.6}~,~~~~~~~r\leq R_* \\ \\
M_*\left(\frac{r}{R_*}\right)^{1.0}~,~~~~~~~r> R_*
\end{array}
\right. \label{mass_star}
\end{equation}
with a total stellar mass $M_*=0.88\times 10^6$ M$_{\odot}$ and a
size $R_*=0.3878$ pc.

As far as the mass profile of the DM concentration is concerned,
\citet{hallgondolo} have assumed a mass distribution of the form
\begin{equation}
M_{DM}(<r)=\left\{
\begin{array}{ll}
M_{DM}\left(\frac{r}{R_{DM}}\right)^{3-\alpha}~,~~~~~~~r\leq R_{DM} \\ \\
M_{DM},~~~~~~~~~~~~~~~~~~~~~~r> R_{DM}
\end{array}
\right. \label{mass_dm}
\end{equation}
$M_{DM}$ and $R_{DM}$ being the total amount of DM in the form of
WIMPs and the radius of the spherical mass distribution,
respectively.

%A likelihood analysis has allowed to estimate for
%the DM mass the value $M_{DM}\simeq 10^5$ M$_{\odot}$ while the DM
%sphere size results to be in the range $10^{-4}-1$ pc.
Hall and Gondolo \cite{hallgondolo} discussed limits on DM mass
around the black hole at the Galactic Center. It is clear that
present observations of stars around the Galactic Center do not
exclude the existence of a DM sphere with mass $\simeq 4 \times
10^6 M_{\odot}$, well contained within the orbits of the known
stars, if its radius $R_{DM}$ is $\lesssim 2\times 10^{-4} $ pc
(the periastron distance of the S16 star in the more recent
analysis \cite{Ghez_2005}). However, if one considers a DM sphere
with larger radius, the corresponding upper value for $M_{DM}$
decreases (although it tends again to increase for extremely
extended DM configurations with $R_{DM}\gg 10$ pc). In the
following, we will assume for definiteness a DM mass $M_{DM} \sim
2 \times 10^5 M_\odot$, that is the upper value for the DM sphere
in \cite{hallgondolo} within an acceptable confidence level in the
range $10^{-3}-10^{-2}$ pc for $R_{DM}$. As it will be clear in
the following, we emphasize that even a such small value for the
DM mass (that is about only 5\% of the standard estimate $3.67\pm
0.19\times 10^6~M_{\odot}$ for the dark mass  at the Galactic
Center \cite{Ghez_2005}) may give some observational signatures.

Evaluating the S2 apoastron shift \footnote{We want to note that
the periastron and apoastron shifts $\Delta\Phi$ as seen from the
orbit center have the same value whereas they have different
values as seen from Earth (see Eq. (\ref{deltaphiE})). When we are
comparing our results with orbit reconstruction from observations
we refer to the apoastron shift as seen from Earth.} as a function
of $R_{DM}$, one can further constrain the DM sphere radius since
even now we can say that there is no evidence for negative
apoastron shift for the S2 star orbit at the level of about 10 mas
\cite{Genzel_2003}. In addition, since at present the precision of
the S2 orbit reconstruction is about 1~mas, we can say that even
without future upgrades of the observational facilities and simply
monitoring the S2 orbit, it will be possible within about 15 years
to get much more severe constraints on $R_{DM}$.

Moreover, observational facilities will allow in the next future
to monitor faint infrared objects at the astrometric precision of
about 10 $\mu$as \cite{Eisenhauer_2005} and, in this case,
previous estimates will be sensibly improved since it is naturally
expected to monitor eccentric orbits for faint infrared stars
closer to the Galactic Center with respect to the S2 star.

\begin{figure}
\begin{center}
\includegraphics[scale=0.60]{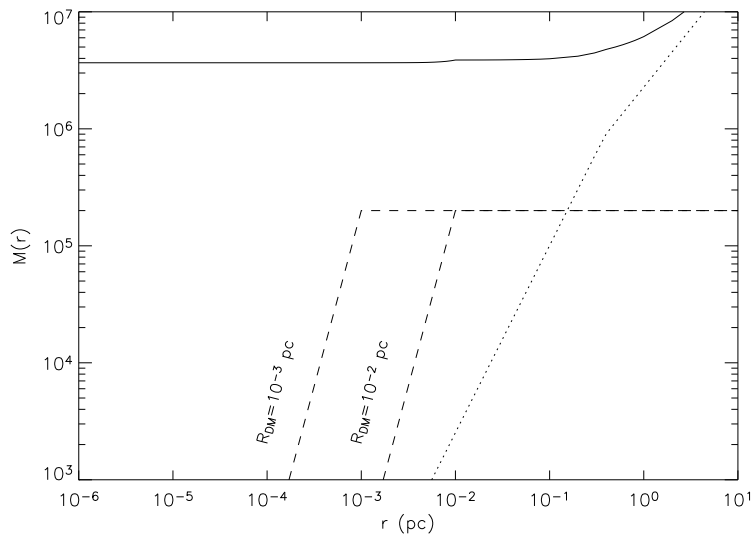}\qquad
\end{center}
\caption{The mass $M(<r)$ obtained in \cite{Ghez_2003} from
observations of stars at the Galactic Center is shown (solid
line). The dotted line represents the stellar mass profile as
given in Eq.~(\ref{mass_star}), while the dashed lines are for DM
spheres with radii $R_{DM}=10^{-3}$ and $10^{-2}$ pc and mass
$M_{DM}\simeq 2\times 10^5$ M$_{\odot}$, that corresponds to some
acceptable estimate for the upper limit of $M_{DM}$ from Figs. 4-6
in \cite{hallgondolo} for $R_{DM}$ in the above range of values.}
\label{mass}
\end{figure}

In Fig.~\ref{mass}, the mass profile $M(<r)$ \cite{Ghez_2003}
obtained by using observations of stars nearby the Galactic Center
is shown (solid line). The dotted line represents the stellar mass
profile as given in Eq.~(\ref{mass_star}), while the dashed lines
are for DM spheres with mass $M_{DM}\simeq 2\times 10^5$
M$_{\odot}$ and radii $R_{DM}=10^{-3}$ and $10^{-2}$ pc,
respectively.

%%%%%%%%%%%%%%%%%%%%%%%%%%%%%%%%%%%%%%%%%%%%%%%%%%%%%%%%%%

% figure Density
%%%%%%%%%%%%%%%%%%%%%%%%%%%%%%%%%%%%%%%%%%%%%%%%%%%%%%%%%

\begin{figure}
\begin{center}
\includegraphics[scale=0.60]{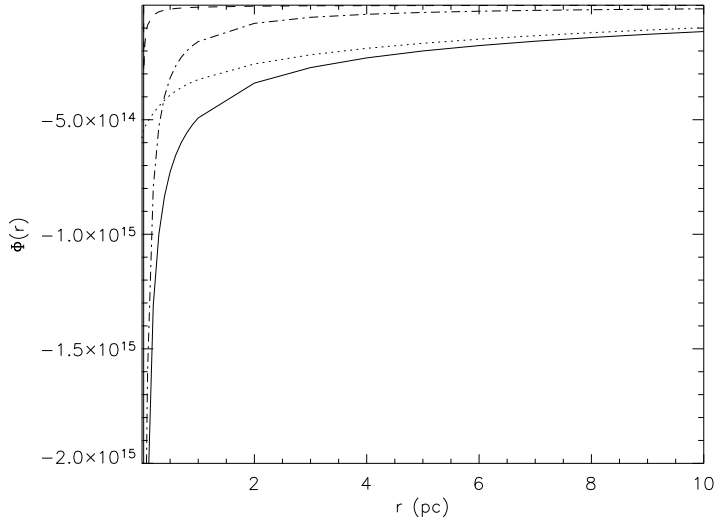}\qquad
\end{center}
\caption{The gravitational potential  $\Phi(r)$ (solid line) in
cgs units as a function of the galactocentric distance $r$ as due
to the mass $M(r)$ in Eq.(\ref{totalmass}) is shown. For
comparison, also the gravitational potentials due to the single
mass components, i.e. black hole (dashed line), stellar cluster
(dot-dashed line) and DM (dotted line), are also given. Here we
assume that DM mass $M_{DM}\simeq 2\times 10^5$ M$_{\odot}$ and
radius $R_{DM}=10^{-3}$ pc.} \label{potential}
\end{figure}

In the following section, we study the motion of stars as a
consequence of the gravitational potential $\Phi(r)$ due the mass
profile given in Eq.~(\ref{totalmass}). As usual, the
gravitational potential can be evaluated as
\begin{equation}
\Phi(r) = -G \int_r^{\infty} \frac{M(r')}{r'^2}~dr'~.
\label{gravitationalpotential}
\end{equation}
For convenience, in Fig.~\ref{potential} the gravitational
potential due to the total mass (solid line) contained within $r$
is given as  function of the galactocentric distance. For
comparison, the contributions due to the single mass components,
i.e. central black hole, stellar cluster and DM sphere, are also
shown.

%%%%%%%%%%%%%%%%%%%%%%%%%%%%%%%%%%%%%%%%%%%%%%%%%%%%%%%%%%

% figure Density

%%%%%%%%%%%%%%%%%%%%%%%%%%%%%%%%%%%%%%%%%%%%%%%%%%%%%%%%%

\section{Apoastron Shift Constraints}

According to GR, the motion of a test particle can be fully
described by solving the geodesic equations. Under the assumption
that the matter distribution is static and pressureless, the
equations of motion at the first post-Newtonian (PN) approximation
become (see e.g. \cite{Fock,Weinberg72,Rubilar01})
\begin{equation}
\frac{d\textbf{v}}{dt}\simeq-\nabla(\Phi_N +2\Phi_N
^2)+4\textbf{v}(\textbf{v}\cdot \nabla)\Phi_N -v^2\nabla \Phi_N~.
\label{5}
\end{equation}
We note that the PN-approximation is the first relativistic
correction from which the apoastron advance phenomenon arises
\footnote{Obviously, to take into account gravitational radiation
reaction, at least a 2.5 PN approximation is needed  (see e.g.
\cite{Grishchuk}).}. In the case of the S2 star, the apoastron
shift as seen from Earth (from Eq. (\ref{deltaphiE})) due to the
presence of a central black hole is about 1~mas, therefore not
directly detectable at present since the available precision in
the apoastron shift is about 10 mas (but it will become about 1
mas in 10--15 years even without considering possible
technological improvements). It is also evident that higher order
relativistic corrections to the S2 apoastron shift are even
smaller and therefore may be neglected at present, although they
may become important in the future.

As it will be discussed below, the Newtonian effect due to the
existence of a sufficiently extended DM sphere around the black
hole may cause a apoastron shift in the opposite direction with
respect to the relativistic advance due to the black hole.
Therefore, we have considered the two effects comparing only the
leading terms.

For the DM distribution at the Galactic Center we follow Eq.
(\ref{mass_dm}) as done by \cite{hallgondolo}. Clearly, if in the
future faint infrared stars (or spots) closer to the black hole
with respect to the S2 star will be monitored
\cite{Eisenhauer_2005}, this simplified model might well not hold
and higher order relativistic corrections may become necessary.

For a spherically symmetric mass distribution (such as that
described above) and for a gravitational potential given by
Eq.~(\ref{gravitationalpotential}), Eq. (\ref{5}) may be rewritten
in the form (see for details \cite{Rubilar01})
\begin{equation}
\frac{d\textbf{v}}{dt}\simeq-\frac{GM(r)}{r^3}\left[\left(1+\frac{4\Phi_N}{c^2}+
\frac{v^2}{c^2}\right)\textbf{r}-\frac{4\textbf{v}(\textbf{v}\cdot\textbf{r})}{c^2}\right]~,
\label{setode}
\end{equation}
$\textbf{r}$ and $\textbf{v}$ being the vector radius of the test
particle with respect to the center of the stellar cluster and the
velocity vector, respectively. Once the initial conditions for the
star distance and velocity are given, the rosetta shaped orbit
followed by a test particle can be found by numerically solving
the set of ordinary differential equations in eq. (\ref{setode}).

\begin{figure*}[htbp]
\vspace{0.2cm}
\begin{center}
$\begin{array}{c@{\hspace{0.1in}}c@{\hspace{0.1in}}c}
\epsfxsize=3.0in \epsfysize=3.0in \epsffile{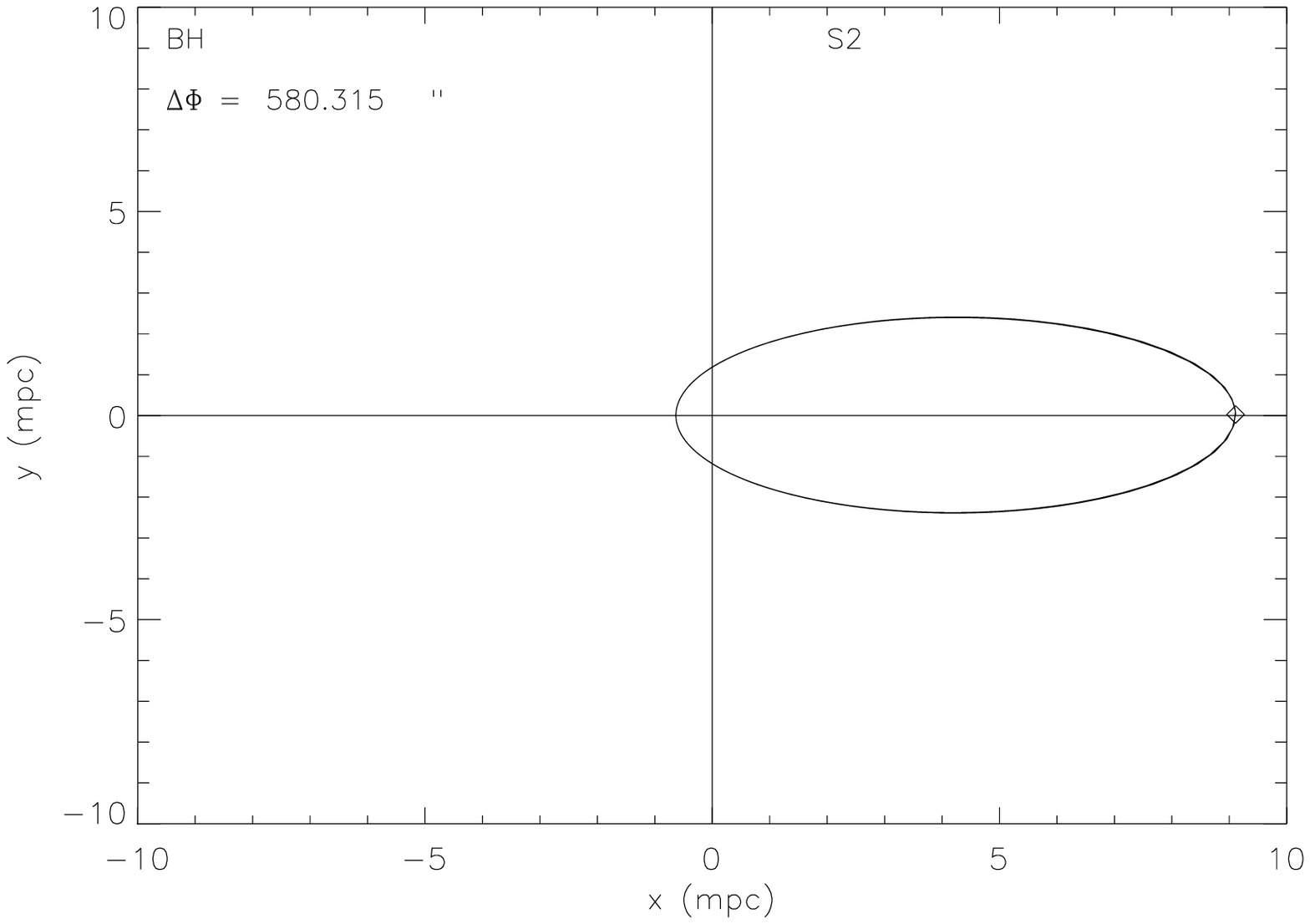} &
\epsfxsize=3.0in \epsfysize=3.0in \epsffile{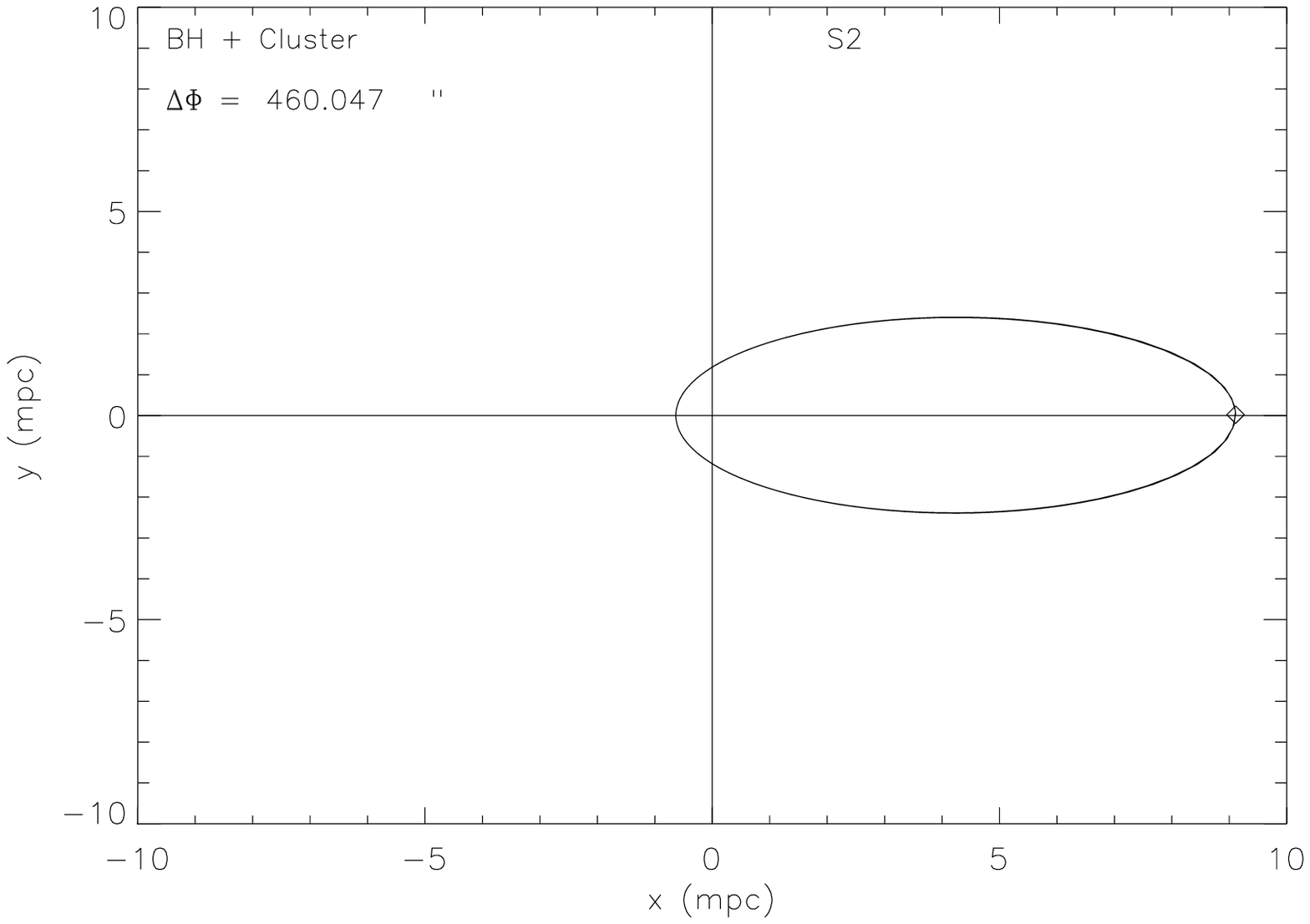} &\\
\epsfxsize=3.0in \epsfysize=3.0in \epsffile{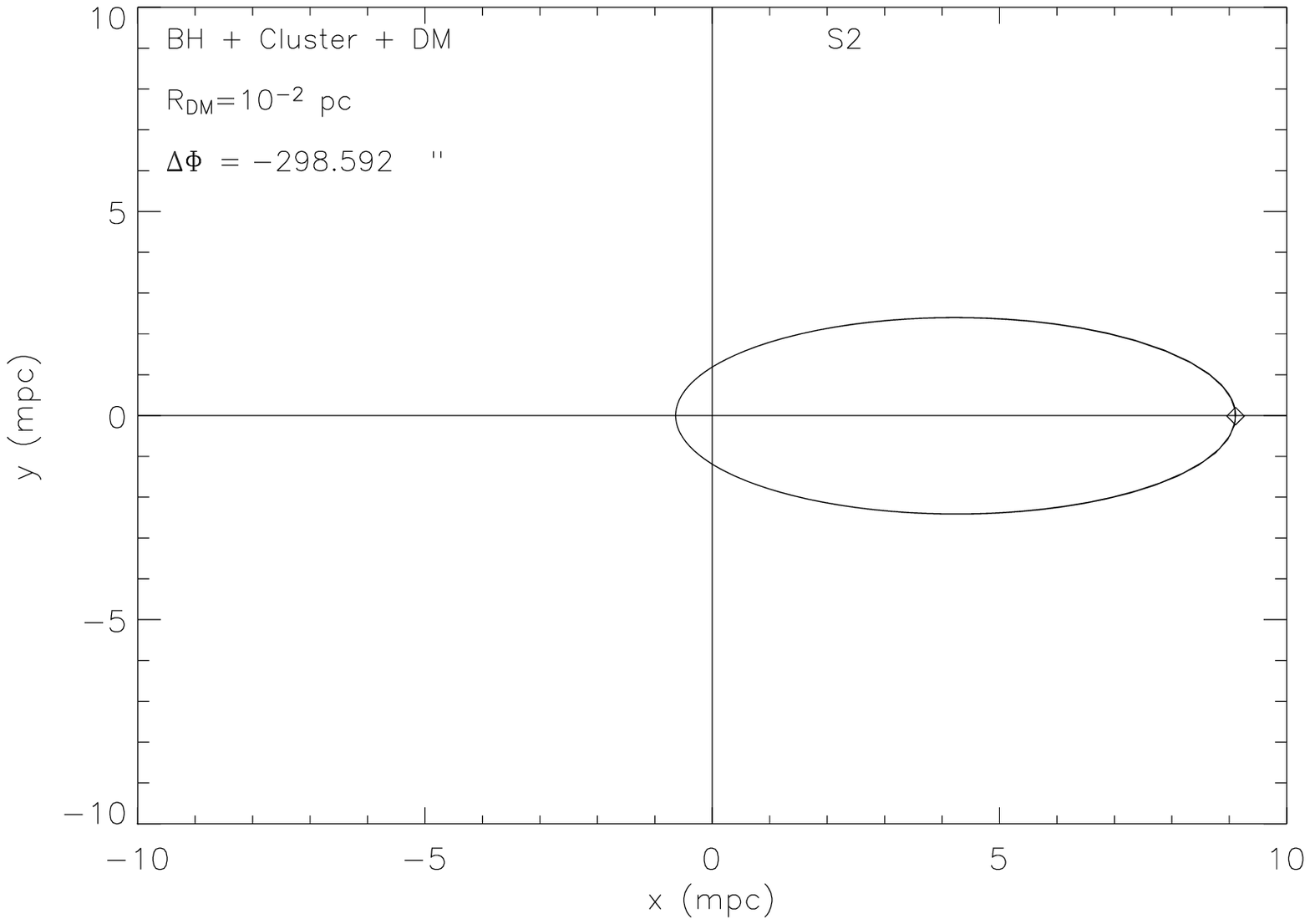} &
\epsfxsize=3.0in \epsfysize=3.0in \epsffile{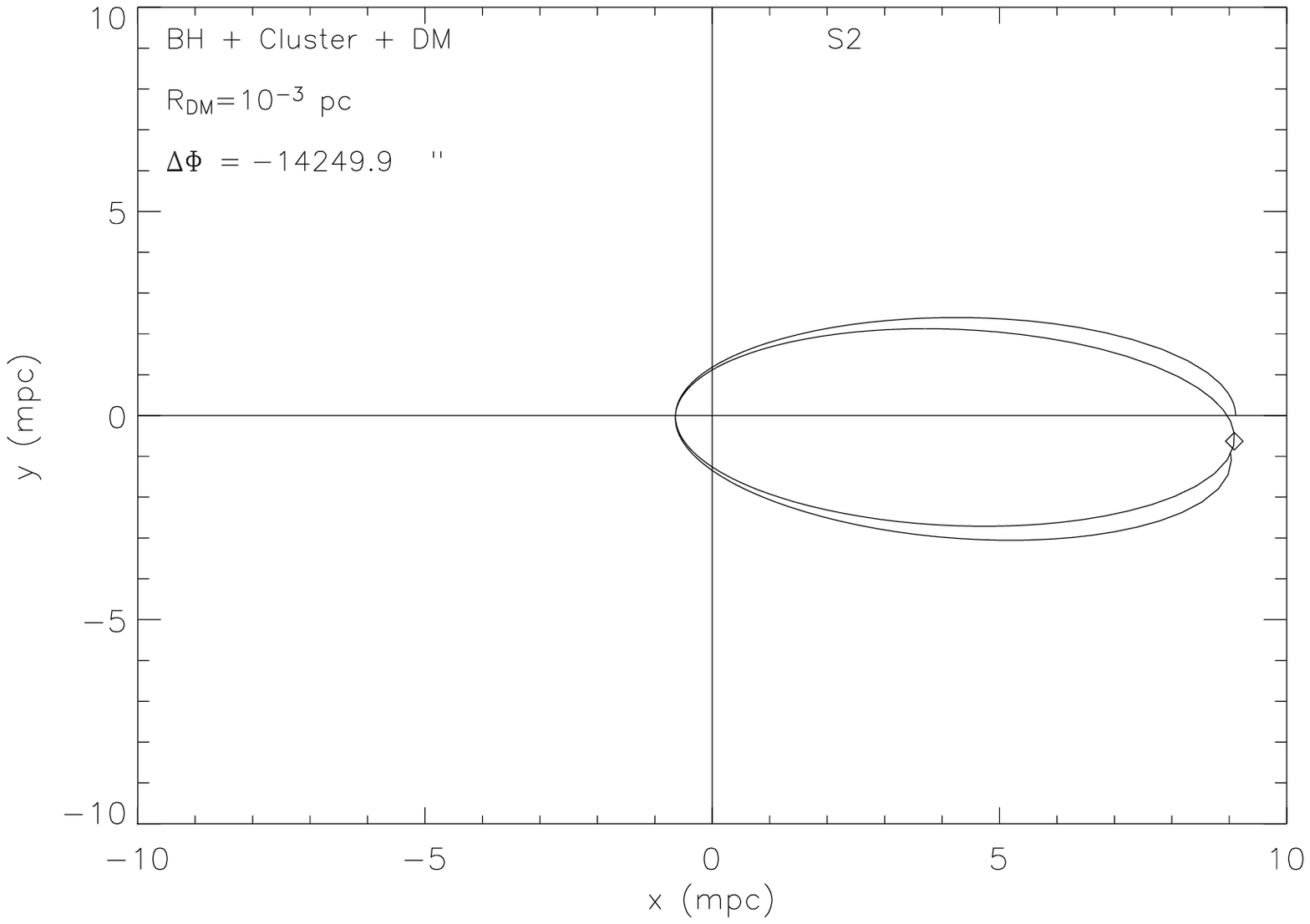} \\
\end{array}$
\end{center}
\caption{PN-orbits for different mass configurations at the
Galactic Center. The S2 star has been considered as a test
particle and its apoastron shift is indicated in each panel as
$\Delta \Phi$ (in arcsec). The top-left panel shows the central
black hole contribution to the S2 shift that amounts to about 580
arcsec. The top-right panels shows the combined contribution of
the black hole and the stellar cluster (taken following eq.
\ref{mass_star}) to the S2 apoastron shift. In the two bottom
panels the contribution due to two different DM mass-density
profiles is added (as derived in eq. \ref{mass_dm}). We assume
that DM mass $M_{DM}\simeq 2\times 10^5$ M$_{\odot}$.}
\label{orbite}
\end{figure*}

In Fig. \ref{orbite}, as an example, assuming that the test
particle orbiting the Galactic Center region is the S2 star, we
show the Post Newtonian orbits obtained by the black hole only,
the black hole plus the stellar cluster and the contribution of
two different DM mass density profiles. In each case the S2 orbit
apoastron shift is given. As one  can see, for selected parameters
for DM and stellar cluster masses and radii the effect of the
stellar cluster is almost negligible while the effect of the DM
distribution is crucial since it enormously overcome the shift due
to the relativistic precession. Moreover, as expected, its
contribution is opposite in sign with respect to that of the black
hole \cite{S2}.

We note that the expected apoastron (or, equivalently, periastron)
shifts (mas/revolution), $\Delta \Phi$ (as seen from the center)
and the corresponding values $\Delta \phi^{\pm} _E$ as seen from
Earth (at the distance $R_0\simeq~8$ kpc from the GC) are related
by
\begin{equation}
\Delta \phi^{\pm}_E = \frac{d(1\pm e)}{R_0} \Delta \Phi,
\label{deltaphiE}
\end{equation}
where with the sign $\pm$ are indicated the shift angles  of the
apoastron (+) and periastron (-), respectively. The S2 star
semi-major axis and eccentricity are $d=919$~AU and $e=0.87$
\cite{Ghez_2005}.

\begin{figure}
\begin{center}
\includegraphics[scale=0.60]{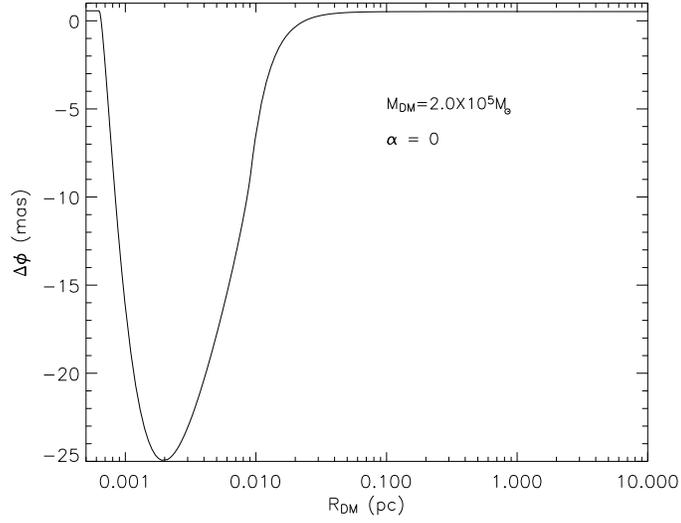}\qquad
\end{center}
\caption{Apoastron shift as a function of the DM radius $R_{DM}$
for $\alpha=0$ and $M_{DM}\simeq 2\times 10^5$ M$_{\odot}$. Taking
into account present day precision for the apoastron shift
measurements (about 10 mas) one can say that DM radii $R_{DM}$ in
the range $8\times 10^{-4}-10^{-2}$ pc are not acceptable.}
\label{shift_alpha0}
\end{figure}

\begin{figure}
\begin{center}
\includegraphics[scale=0.60]{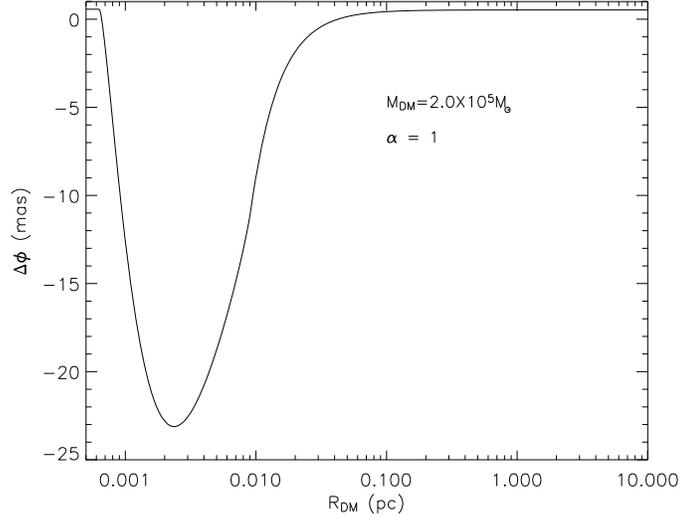}\qquad
\end{center}
\caption{The same as in Fig. \ref{shift_alpha0}  for $\alpha=1$
and $M_{DM}\simeq 2\times 10^5$ M$_{\odot}$. As in the previous
case one can say that the S2 apoastron shift put severe limits on
the DM mass radii that are not acceptable in the range $9\times
10^{-4}-10^{-2}$ pc.} \label{shift_alpha1}
\end{figure}

\begin{figure}
\begin{center}
\includegraphics[scale=0.60]{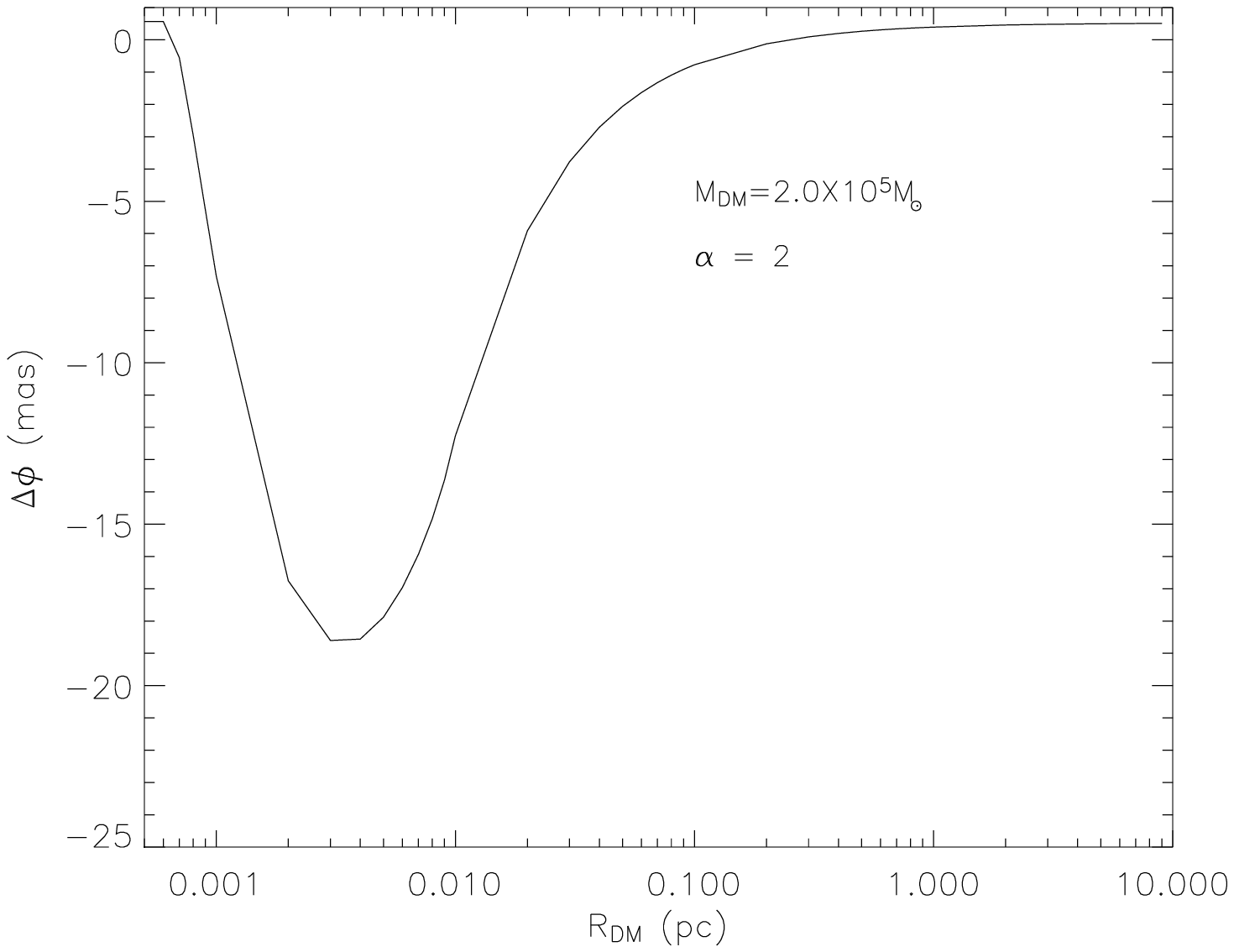}\qquad
\end{center}
\caption{The same as in Fig. \ref{shift_alpha0}  for $\alpha=2$
and $M_{DM}\simeq 2\times 10^5$ M$_{\odot}$. As in the previous
case one can say that the upper limit to the S2 apoastron shift
allows to  constrain the DM radius to be out the range $1.0\times
10^{-3}-1.1\times 10^{-2}$ pc.} \label{shift_alpha2}
\end{figure}

In Fig. \ref{shift_alpha0}, the S2 apoastron shift as a function
of the DM distribution size $R_{DM}$ is given for $\alpha=0$ and
$M_{DM}\simeq 2\times 10^5$ M$_{\odot}$. Taking into account that
the present day precision for the apoastron shift measurements is
of about 10 mas, one can say that the S2 apoastron shift cannot be
larger than 10 mas. Therefore, any  DM configuration that gives a
total S2 apoastron shift larger than 10 mas (in the opposite
direction due to the DM sphere) is excluded.
%Then, Fig.
%\ref{shift_alpha0} may allow to strongly constrain the DM size not
%to be in the range between $1.2\times 10^{-3}$ pc and $1.1\times
%10^{-2}$ pc.
The same analysis is shown in Figs. \ref{shift_alpha1} and
\ref{shift_alpha2} for two different values of the DM mass
distribution slope, i.e. $\alpha=1$ and $\alpha=2$, respectively.
In any case, we have calculated the apoastron shift for the S2
star orbit assuming a total DM mass $M_{DM}\simeq 2\times 10^5$
M$_{\odot}$. As one can see by inspecting Figs.
\ref{shift_alpha0}-\ref{shift_alpha2}, the upper limit of about 10
mas on the S2 apoastron shift may allow to conclude that DM radii
in the range about $10^{-3}-10^{-2}$ pc are excluded by present
observations.

We notice that the results of the present analysis allows to
further constrain the results of the Hall and Gondolo paper
\cite{hallgondolo} who have concluded that if  the DM sphere
radius is in the range  $10^{-3}-1$ pc, configurations with DM
mass up to $M_{DM}=2\times 10^5~M_{\odot}$ are acceptable. The
present analysis shows that DM configurations of the same mass are
acceptable only for $R_{DM}$ out the range between
$10^{-3}-10^{-2}$ pc, almost irrespectively of the $\alpha$ value.

%%%%%%%%%%%%%%%%%%%%%%%%%%%%%%%%%%%%%%%%%%%%%%%%%%% ARRAY DI FIGURE
%%%%%%%%%%%%%%%%%%%%%%%%%%%%%%%%%%%%%%%%%%%%%%%%%%%%%%%%%%%%%%%%%%

\section{\label{discussion} Conclusions}

In this paper we have considered the constraints that the upper
limit (presently of about 10 mas) of the S2 apoastron shift may
put on the DM configurations at the galactic center considered by
Hall and Gondolo \cite{hallgondolo}.

When (in about 10-15 years, even without considering improvements
in observational facilities) the precision of S2 apoastron shift
will be about 1~mas (that is equal to the present accuracy in the
S2 orbit reconstruction) our analysis will allow to further
constrain the DM distribution parameters. In particular, the
asymmetric shape of the curves in Figs.
\ref{shift_alpha0}-\ref{shift_alpha2} imply that any improvement
in the apoastron shift measurements will allow to extend the
forbidden region especially for the upper limit for $R_{DM}$.

In this context, future facilities for astrometric measurements at
a level 10 $\mu$as of faint infrared stars will be extremely
useful \cite{Eisenhauer_2005} and they give a chance to put even
more severe constraints on DM distribution.

In addition, it is also expected to detect faint infrared stars or
even hot spots \cite{Genzel_2007} orbiting the Galactic Center. In
this case, consideration of higher order relativistic corrections
for an adequate analysis of the stellar orbital motion have to be
taken into account.

In our considerations we adopted simple analytical expression and
reliable values for $R_{DM}$ and $M_{DM}$ parameters following
\cite{hallgondolo} just to illustrate the relevance of the
apoastron shift phenomenon in constraining the DM mass
distribution at the Galactic Center. If other models for the DM
distributions are considered (see, for instance
\cite{Merritt_2007} and references therein) the qualitative
aspects of the problem are preserved although, of course,
quantitative results on apoastron shifts may be different.

\begin{acknowledgments}

AFZ is grateful to Dipartimento di Fisica Universita di Lecce and
INFN, Sezione di Lecce where part of this work was carried out and
to the National Natural Science Foundation of China (Grant \#
10233050) and National Basic Research Program of China
(2006CB806300) for partial financial support. We would like to
thank an anonymous referee for constructive remarks.
\end{acknowledgments}

\end{document}